\documentclass[preprint, a4paper, 12pt]{elsarticle}
\usepackage{amsmath}
\usepackage{amssymb}
\usepackage{color}
\usepackage{dsfont}
\usepackage[utf8]{inputenc}
\usepackage{graphicx}
\usepackage[left=2.2cm, right=2.2cm, bottom=3cm, top=2.5cm]{geometry}
\usepackage{microtype}
\usepackage{natbib}
\usepackage{pdflscape}
\usepackage{gensymb} 

\journal{Computers and Geosciences}

\setcitestyle{authoryear,open={(},close={)}}

\definecolor{orange}{rgb}{1, 0.5, 0}
\definecolor{green}{rgb}{0, 0.5, 0}

\newcommand{\hypers}{\boldsymbol{\alpha}}
\newcommand{\params}{\boldsymbol{\theta}}
\newcommand{\data}{\boldsymbol{d}}
\newcommand{\info}{I}
\newcommand{\x}{x}

\newcommand{\mnras}{Monthly Notices of the Royal Astronomical Society}
\newcommand{\changed}{}   
\newcommand{\change}{} 

\begin{document}

\begin{frontmatter}

\title{AMORPH: A statistical program for characterizing amorphous materials by X-ray diffraction$^1$\footnotetext{\changed $^1$MR conceived the project, developed geologic applications, and drafted the manuscript with BB. BB developed the statistical model, wrote the program code, and co-wrote the manuscript.}}

\author[rowe]{Michael C. Rowe\corref{cor1}}
\ead{michael.rowe@auckland.ac.nz}
\cortext[cor1]{Corresponding author}
\address[rowe]{School of Environment, The University of Auckland, Private Bag 92019, Auckland 1142, New Zealand}

\author[brewer]{Brendon J. Brewer}
\ead{bj.brewer@auckland.ac.nz}
\address[brewer]{Department of Statistics, The University of Auckland, Private Bag 92019, Auckland 1142, New Zealand}

\begin{abstract}
AMORPH utilizes a new Bayesian statistical approach to interpreting X-ray diffraction results of {\changed samples with both crystalline and amorphous components}. AMORPH fits X-ray diffraction patterns with a mixture of {\changed narrow and wide components},
{\changed simultaneously inferring} all of the model parameters and {\changed quantifying} their
uncertainties.
The program simulates background patterns previously {\changed applied manually}, providing reproducible results, and significantly reducing inter- and intra-user biases. This approach allows for the quantification of amorphous and crystalline materials and for the characterization of the amorphous component, including {\changed properties such as the} centre of mass, width, skewness, and nongaussianity of the amorphous component. Results demonstrate the applicability of this program for calculating amorphous contents of volcanic materials and independently modeling their properties in compositionally variable materials.
\end{abstract}


\end{frontmatter}


\setlength{\parskip}{0.7em}

\section{Introduction}
Quantification and characterization of amorphous materials is an important area of research in
terrestrial and planetary sciences \citep[e.g.,][]{schmidt2009, dehouck2014, wall2014, morris2016, zorn2018}.
Powder X-ray diffraction (XRD), although primarily a tool for characterization of crystalline materials,
also provides a means to investigate amorphous materials \citep[e.g.,][]{rowe2012, achilles2013}. X-ray {\changed amorphous materials lack long-range crystallographic order and produce broad humps with low intensities in XRD patterns. X-ray} diffraction of rocks/soils containing amorphous materials (e.g. volcanic
glass, amorphous silica, and organics) produces a diffraction pattern which is a combination of the
crystalline and amorphous phases. Recent studies have debated the nature of amorphous materials in
planetary sciences \citep[e.g.][]{schmidt2009, dehouck2014}, and a growing number of
applications to terrestrial volcanics have presented an avenue of expansion for XRD methodologies
for amorphous material analysis \citep{ellis2015, kanakiya2017, zorn2018}.
Here we present a new statistical approach and computer program, AMORPH, which utilizes Bayesian inference to determine
the crystallinity, and other characteristics, of partially amorphous materials from X-ray diffraction patterns, as well as the uncertainty in these quantities.

Several techniques currently exist for the quantification of amorphous materials in heterogeneous (amorphous + crystalline) materials. However, these
each present their own limitations. The two most commonly used approaches in recent literature
either rely on a calibration curve for different crystallinities \citep[calibration method;][]{ rowe2012} or a combined Rietveld-Reference Intensity Ratio (RIR) method \citep{gualtieri1996, gualtieri2000}. While the combined Rietveld-RIR procedure allows for quantification of both the amorphous 
and crystalline materials, for proper characterization it requires a library of pure components,
spiked with a reference material (e.g. corundum). Sample spiking presents additional complications
as it reduces the observed intensity of the amorphous material thus reducing the accuracy of
pattern matching to determine abundances. An Excel-based adaptation of this pattern matching
technique \citep[FULLPAT;][]{chipera2002} is routinely utilized for characterizing remotely acquired X-ray
diffraction patterns, such as those {\changed measured by the CheMin instrument on} the Mars Science Laboratory (MSL) Curiosity rover \citep{blake2012, bish2013}. The problem faced by RIR approaches is they are based on modeling of an observed 
diffraction pattern compared to library measurements, and therefore do not allow an independent
assessment of the characteristics of the amorphous material. {\changed Furthermore, natural X-ray amorphous phases have a wide variety of compositions such that the library measurements may not include the amorphous phase(s) present in the sample.} In comparison, the calibration method
focuses solely on the quantification of the amorphous material. {\changed This method utilizes} the integrated counts
associated with the amorphous and crystalline componentry: 

\begin{align}
C_{\rm m} \% = [C_{\rm pa}]/[A_{\rm pa} + C_{\rm pa}] \times 100\label{eqn:definition}
\end{align}

where $C_{\rm m}$ is the measured crystallinity, and $C_{\rm pa}$ and $A_{\rm pa}$ are the integrated peak areas for the
crystalline and amorphous components, respectively (see Figure 5 of \citet{wall2014}). 
While its simplicity has made it advantageous, the calibration methodology has been limited in that
the quality of the results largely depend on the ability of the user to manually assign
backgrounds for count integration, resulting in potentially large inter-user variations. Although the
point of the calibration curve is to reduce the inter-user variation, determinations still tend to
be associated with moderate user error (3--5 vol \%). As every user must independently practice
background fittings using the calibration method, the results ultimately are only as good as the
user’s ability to manually {\changed fit} the data. This technique, while not constrained by a
necessary library of diffraction patterns, also does not characterize the XRD properties of the
amorphous material. This study presents a new approach using Bayesian inference to automate
background determination for the calibration method, removing inter-user and intra-user biases in
the determination of crystallinity and peak characteristics for amorphous/partially amorphous
materials, regardless of what they are. 

This paper is structured as follows. In Section~\ref{sec:bayes}, we outline the Bayesian
framework of inference in general terms. In Section~\ref{sec:model} we describe the
specific modelling assumptions made by AMORPH, and
Section~\ref{sec:priors} lists the probabilistic assumptions (i.e., the prior distributions)
{\changed assumed by} the program. Section~\ref{sec:applications} presents some
example applications and comparisons with standard techniques, and finally we conclude
in Section~\ref{sec:conclusions}. Basic instructions for installation and use of AMORPH are given in~\ref{sec:program}.

\section{Bayesian inference}\label{sec:bayes}
AMORPH is founded upon Bayesian inference
\citep{gregory2005bayesian, o2004kendall, sivia2006data}, where
probability distributions are used to model
states of uncertainty about the values of unknown quantities.
Bayesian inference is often implemented computationally with
Markov Chain Monte Carlo (MCMC) methods \citep{mackay2003}, and AMORPH is no exception.
Loosely speaking, the goal is to fit a dataset with a model by exploring
those regions of the model's parameter space that are consistent with the
data. The model assumptions are described in detail in
Section~\ref{sec:model}, and the computation is performed with
Diffusive Nested Sampling {\changed \citep[an advanced MCMC method][]{dns}
as implemented in the software package DNest4 \citep{dnest4}.
This method is robust to complicated posterior distributions.

There have been many approaches to
problems of separating background from compact signals
when noise is present, such as the famous CLEAN algorithm \citep{clean},
through to more explicitly Bayesian approaches
\citep[e.g.][]{padayachee, massinf, object, hobson, worpel}.
AMORPH can be viewed as another
case of this, with assumptions customized to this specific application.}

Typically, one starts with the `prior distribution'
for unknown quantities (`parameters') $\params$, written $p(\params | \info)$
(the probability distribution for $\params$ {\em given} prior information
$\info$). Throughout this paper, the bold $\params$ is used to denote a
collection of unknown parameters, whereas $2\theta$ (with the factor of two, and the
non-bold $\theta$) is later used to
refer to the incident angle plus reflected angle, which is the
standard notation in X-ray diffraction. 

The prior describes an initial state of uncertainty about the
parameter values, and is often a wide probability distribution.
The prior is supplemented with $p(\data | \params, \info)$, sometimes
called a `sampling distribution',
which describes the uncertainty about the data which we {\em would} have,
if we knew the parameters (and the prior information), as a function of
the parameters. {\changed The sampling distribution} describes uncertainty about what data will be observed,
but encodes some knowledge about the fact that the data have something to do with the
parameters $\params$. For example, $p(\data | \params, \info)$ often encodes
the assumption that the data has noise in it, and the specific noise values
are unknown a priori.

The
product of these two distributions is the joint prior
\begin{align}
p(\params, \data | \info) &= p(\params | \info)p(\data | \params, \info)
\end{align}
which describes uncertainty about the value of both the parameters and the
dataset. Once a specific observed dataset $\data_{\rm obs}$ is known, knowledge about
$\params$ is updated from the prior $p(\params | \info)$ to the posterior
\begin{align}
p(\params | \data_{\rm obs}, \info) &\propto
    p(\params | \info)p(\data_{\rm obs} | \params, \info).
\end{align}
which takes the specific data $\data_{\rm obs}$ into account {\changed via conditioning}.
With the specific dataset,
$p(\data_{\rm obs} | \params,\info)$ becomes a function of the
unknown parameters $\params$ only and is called the {\em likelihood function}.
The posterior distribution is therefore proportional to the prior distribution
times the likelihood function.

When $\params$ is a large collection of parameters, the posterior distribution
is a probability distribution over a high-dimensional parameter space. To
represent it computationally, Markov Chain Monte Carlo (MCMC) methods are
often used to generate samples from the parameter space, according to the
posterior distribution. The output is a collection of plausible values of the
parameters which can be used to approximate posterior probabilities.
For example, the probability some parameter $\phi$ is greater than 3.4
is approximately the proportion of the posterior samples that have
$\phi > 3.4$. While many data analysis procedures work by finding the
best fitting values of the parameters (i.e., a maximum likelihood estimate),
Bayesian inference via MCMC involves {\changed exploring a range of plausible values
for the parameters. Since AMORPH uses Diffusive Nested Sampling, it also calculates
the `marginal likelihood' or `evidence' quantity $p(\data | \info)$, which
quantifies how well the model fits the data overall, averaged over all
possible values of its parameters. This allows straightforward model comparison
with any alternatives that may be proposed in the future
\citep{skilling2006nested}.

The computational cost of running AMORPH
varies, as the posterior distribution depends on the data.
The size of the dataset is important, of course.
Another important factor is the number of MCMC samples desired.
To quantify the uncertainty about a few scalar parameters, about
100 posterior samples is usually sufficient. Indeed, \citet[][p. 380]{mackay2003} suggests
that only about twelve are needed for the uncertainty about a single
quantity!
However, if one wants to produce
smooth histograms or look at details of the posterior distribution,
many more samples are required.

On a modern desktop PC and using aggressive numerical settings
(see~\ref{sec:program} for details),
some useful posterior samples can be obtained in about
15 minutes when the dataset contains about 1500 points.
Typical analyses take approximately 5 hours to fully complete and yield
a few hundred posterior samples.
However, the complete runs presented in this paper took
about a day each to run, as we ran with more conservative settings
to ensure the number of posterior samples was about 1000 for each dataset.}

\section{The model curve}\label{sec:model}

\begin{figure}[!ht]
\centering
\includegraphics[width=.7\textwidth]{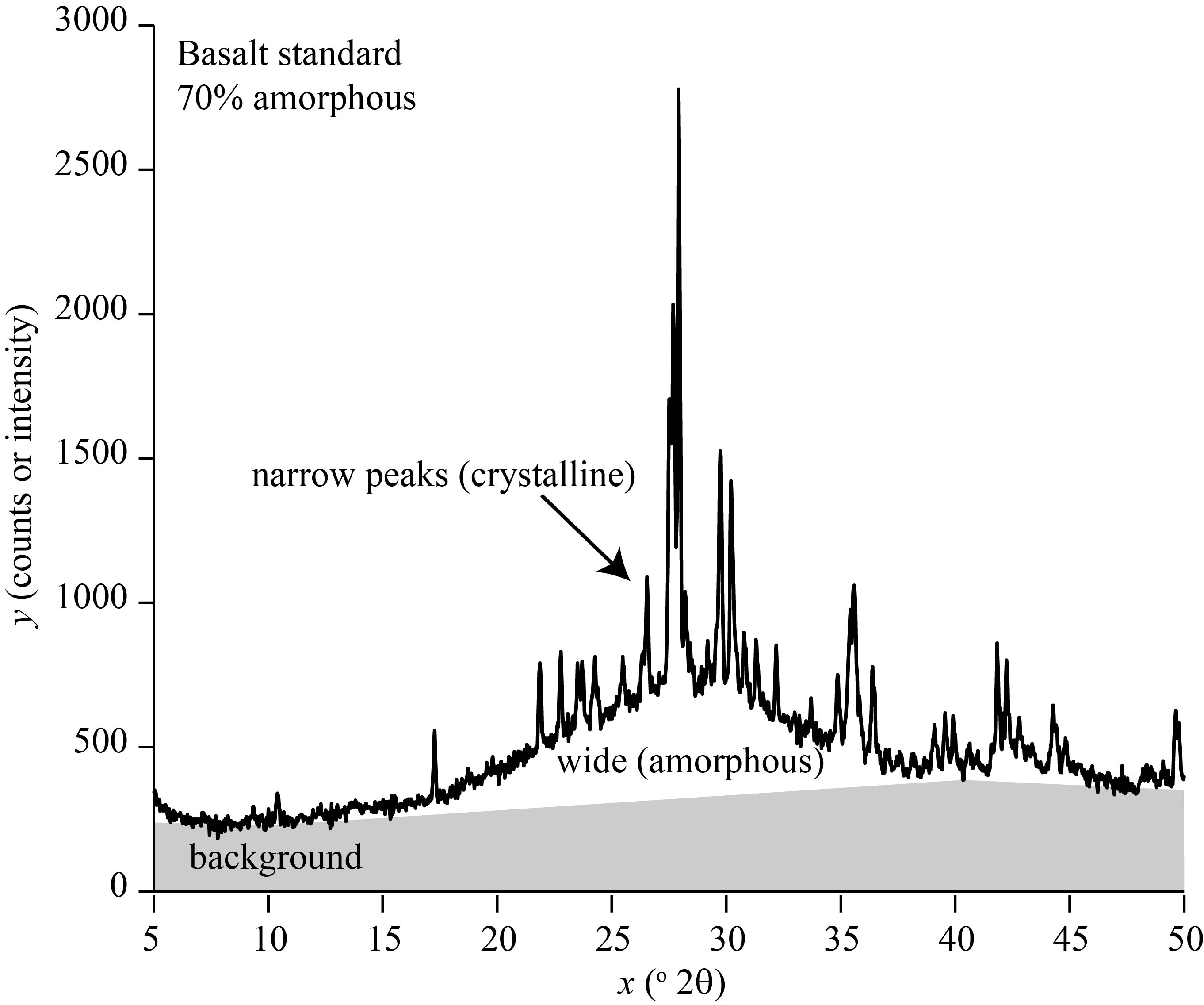}
\caption{\it Powder X-ray diffraction results for a basalt standard with 70\% amorphous material. Annotation corresponds to the model components: background, amorphous (wide) component, and crystalline component (narrow peaks).\label{fig:example_data2}}
\end{figure}

AMORPH fits the data with a model curve defined over some range from
$x_{\rm min}$ to $x_{\rm max}$ (Fig.~\ref{fig:example_data2}). Throughout this section, we use
$x$ as a synonym for $2\theta$, since the problem is basically a curve-fitting
problem where $x$ and $y$ are typically used.

The model curve is a function $y = f(\x; \params)$,
which is assumed to be a sum of the following components identified in Figure~\ref{fig:example_data2}:
\begin{enumerate}
\item A background component, described by parameters $\params_{\rm bg}$
\item An amorphous component, described by parameters $\params_{\rm amorph}$
\item A crystalline component, described by parameters $\params_{\rm xtal}$
\end{enumerate}
The total model function $f_{\rm tot}(x)$ is therefore
\begin{align}
f_{\rm tot}(\x) &= f_{\rm bg}(\x; \params_{\rm bg})
       + f_{\rm amorph}(\x; \params_{\rm amorph})
       + f_{\rm xtal}(\x; \params_{\rm xtal})
\end{align}
and the measured amorphous content ($A_{\rm m}$ or $1-C_{\rm m}$) is defined,
similar to Equation~\ref{eqn:definition}, as the proportion of the area in the
amorphous and crystalline components which is accounted for by the amorphous component:
\begin{align}
{A_{\rm m} = 1-C_{\rm m}} &= \frac{\int_{\x_{\rm min}}^{\x_{\rm max}}
            f_{\rm amorph}(\x; \params_{\rm amorph}) \, d\x}
          {\int_{\x_{\rm min}}^{\x_{\rm max}}
            f_{\rm amorph}(\x; \params_{\rm amorph}) \, dx
            + \int_{\x_{\rm min}}^{\x_{\rm max}} f_{\rm xtal}(\x; \params_{\rm xtal}) \, d\x} \label{eqn:crystallinity}
\end{align}

From each sample from the posterior distribution for all of the parameters,
it is straightforward to compute the proportion of amorphous material, and hence
calculate the marginal posterior distribution of the amorphous content.
In the following subsections, we describe the specific parameterizations of the
three parts of the model curve, that is, the functional forms of
$f_{\rm bg}$, $f_{\rm amorph}$, and $f_{\rm xtal}$.

\subsection{The background component}
The background component is assumed to be piecewise linear, with constant
slope
{between $\x=\x_{\rm min}$ and $\x=10$\degree, between $\x=10$\degree~and $\x=40$\degree,
and between $\x=40$\degree~and $\x=\x_{\rm max}$}. These values are selected to maintain a linear background over the interval from 10--40\degree, the predominant range of amorphous materials {\changed (8.8 -- 2.25 \AA)} analyzed using a Cu K$\alpha$ x-ray source {\changed (Table~\ref{tab:anode})}, while still allowing for some flexibility in the structure of the background outside this range.

\begin{figure}[!ht]
\centering
\includegraphics[width=0.6\textwidth]{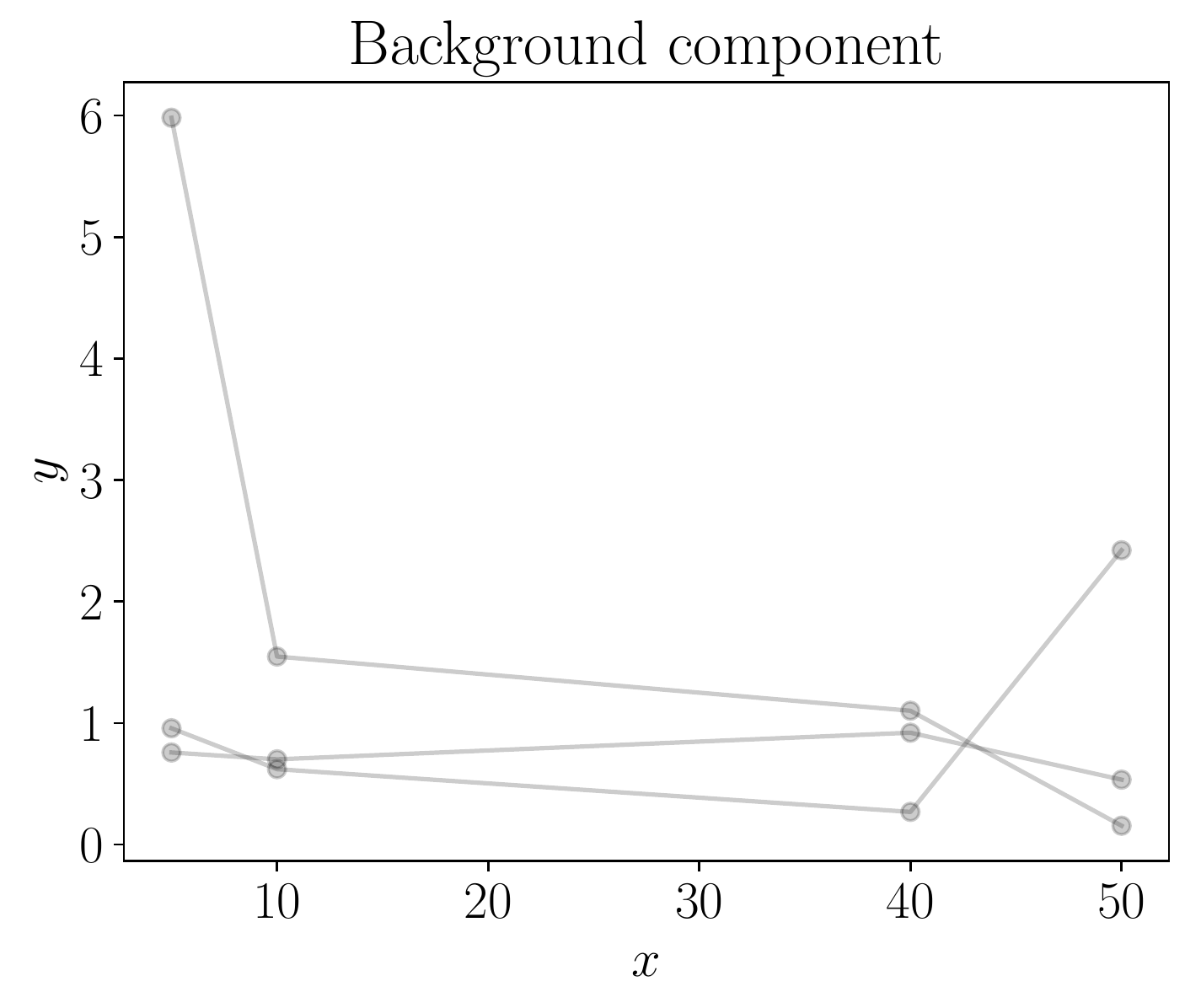}
\caption{\it Three example functions showing possible shapes of the
background component.\label{fig:background}}
\end{figure}

We parameterized the background component using
five parameters. Firstly, there is a mean level $b$, which sets the
typical amplitude of the background component. There are also four parameters
$n_1, ..., n_4$, which determine the offset of four `control points'
from the mean level. The positions of the four control points are then:
\begin{align}
\left(\x_{\rm min}, be^{n_1}\right) \nonumber\\
\left(10, be^{n_2}\right) \nonumber\\
\left(40, be^{n_3}\right) \nonumber\\
\left(\x_{\rm max}, be^{n_4}\right). \nonumber
\end{align}
{\changed Control points may be set by the user,
as it may depend on the X-ray source in use and changes in the required range for the linear background as a function of the nature of the amorphous material (Table~\ref{tab:anode}).} The value of the background component at any other point is found by
linear interpolation of these control points.
Figure~\ref{fig:background} shows three example
background curves generated from this parameterization, with typical
values of the parameters.

\begin{table}
\footnotesize
\centering
{\changed
\begin{tabular}{|llll|}
\hline
{\bf X-ray Anode}      &   {\bf X-ray wavelength}   &  {\bf Control point 2} & {\bf Control point 3}\\
 &   {(\AA)}   &  {\bf(\degree $2\theta$)} & {\bf(\degree $2\theta$)}\\
\hline
Cr & 2.29 & 15.0 & 61.2\\
Fe & 1.94 & 12.7 & 51.1\\
Co & 1.79 & 11.7 & 46.9\\
{\bf Cu} & {\bf 1.54} & {\bf 10.0} & {\bf 40.0}\\
Mo & 0.71 & 4.6 & 18.2\\
Ag & 0.56 & 3.6 & 14.3\\
\hline
\end{tabular}
\caption{\it Typical XRD anodes with recommended background control points based on a predominant 8.8--2.25 {\AA}  amorphous peak.\label{tab:anode}}
} 
\end{table}

\subsection{The crystalline component}
We assume that there is some number $N_{\rm xtal}$ of narrow peaks
{\changed due to crystalline material},
each of which has {\changed the same shape. The shapes of the narrow
peaks are controlled by a single shape parameter $\nu_{\rm xtal}$,
assumed common to all of the narrow peaks. The chosen shape is based
on the Student's $t$-distribution, which allows the peaks to be heavy-tailed
if $\nu_{\rm xtal}$ is low, and to be Lorentzian or Cauchy in shape if
$\nu_{\rm xtal}$ is exactly 1. As $\nu_{\rm xtal}$ is high, the shape
approximates a gaussian.}

The $i$th peak is parameterized
by its central position $X_{i, \rm xtal}$,
{\changed total integrated `mass' $M_{i, \rm xtal}$}, and width $W_{i, \rm xtal}$.
The parameter vector for crystalline component is {\changed
\begin{align}
\params_{\rm xtal} &=
  \{N_{\rm xtal}, \nu_{\rm xtal}, X_{1, \rm xtal}, ..., X_{N, \rm xtal},
    M_{1, \rm xtal}, ..., M_{N, \rm xtal},
    W_{1, \rm xtal}, ..., W_{N, \rm xtal}\}
\end{align} } 
and the total shape contributed by the
crystalline component is a sum of the $N_{\rm xtal}$ {\changed peaks:
\begin{align}
f_{\rm xtal}(\x; \params_{\rm xtal}) &=
	\frac{\Gamma(\frac{\nu_{\rm xtal} + 1}{2})}{\sqrt{\nu_{\rm xtal}\pi}
                 		\Gamma(\frac{\nu_{\rm xtal}}{2})}
    \sum_{i=1}^{N_{\rm xtal}}
    		\frac{M_{i, \rm xtal}}
            	 {W_{i, \rm xtal}}
	\left(1 + \frac{1}{\nu}\left(\frac{x - X_{i, \rm xtal}}{W_{i, \rm xtal}}\right)^2\right)
    			^{-\frac{\nu_{\rm xtal} + 1}{2}}. \label{eqn:t_shape}
\end{align}
The terms inside the sum have the form of the probability density function
of the $t$-distribution, adjusted by shifting according to the central
position and widening by the width, and the total integral is scaled by
the mass parameters. $\Gamma()$ is the gamma function.

For computational efficiency, evaluation of Equation~\ref{eqn:t_shape}
is performed using a lookup table. In addition, we evaluate the peak
shapes at the $x$-position of each data point, rather than integrating
the peak shape across a finite bin width. This is faster but must be
kept in mind when interpreting parameters affected by this assumption,
such as the widths of the narrow peaks. } 

\subsection{The amorphous component}
The amorphous component is also composed of {\changed a sum of peaks, which we chose to be
gaussians. The parameterization is the same as for the crystalline peaks but with a fixed,
high value for the shape parameter.
Apart from the gaussian shape, the only other difference is the prior distributions for the positions, masses, and widths, given in Section~\ref{sec:priors}. In short, the wide gaussians representing the
amorphous component are more likely to be located near the middle of the
dataset, and to be wider.
The constrainted prior for the positions of the wide amorphous
peaks prevents the program from attempting to explain background
variations near the edge of the data using the amorphous component.}

There is some number $N_{\rm amorph}$ of gaussians in the amorphous component.
The $i$th gaussian is parameterized
by its central position $X_{i, \rm amorph}$,
{\changed mass (total integral) $M_{i, \rm amorph}$}, and width $W_{i, \rm amorph}$.
The parameter vector for the wide gaussians is {\changed
\begin{align}
\params_{\rm amorph} =
  \{N_{\rm amorph}, X_{1, \rm amorph}, ..., X_{N, \rm amorph},
    M_{1, \rm amorph}, ..., M_{N, \rm amorph},\\
   \quad\quad W_{1, \rm amorph}, ..., W_{N, \rm amorph}\}
\end{align} } 
and the total shape contributed by the
amorphous component is a sum of the $N_{\rm amorph}$ gaussians: {\changed
\begin{align}
f_{\rm amorph}(\x; \params_{\rm amorph}) &=
	\frac{1}{\sqrt{2\pi}}
    \sum_{i=1}^{N_{\rm amorph}} \frac{M_{i, \rm amorph}}{W_{i, \rm amorph}}
 \exp\left[-\frac{1}{2W_{i, \rm amorph}^2}\left(x - X_{i, \rm amorph}\right)^2\right].
\end{align} } 

\begin{figure}[!ht]
\centering
\includegraphics[scale=0.7]{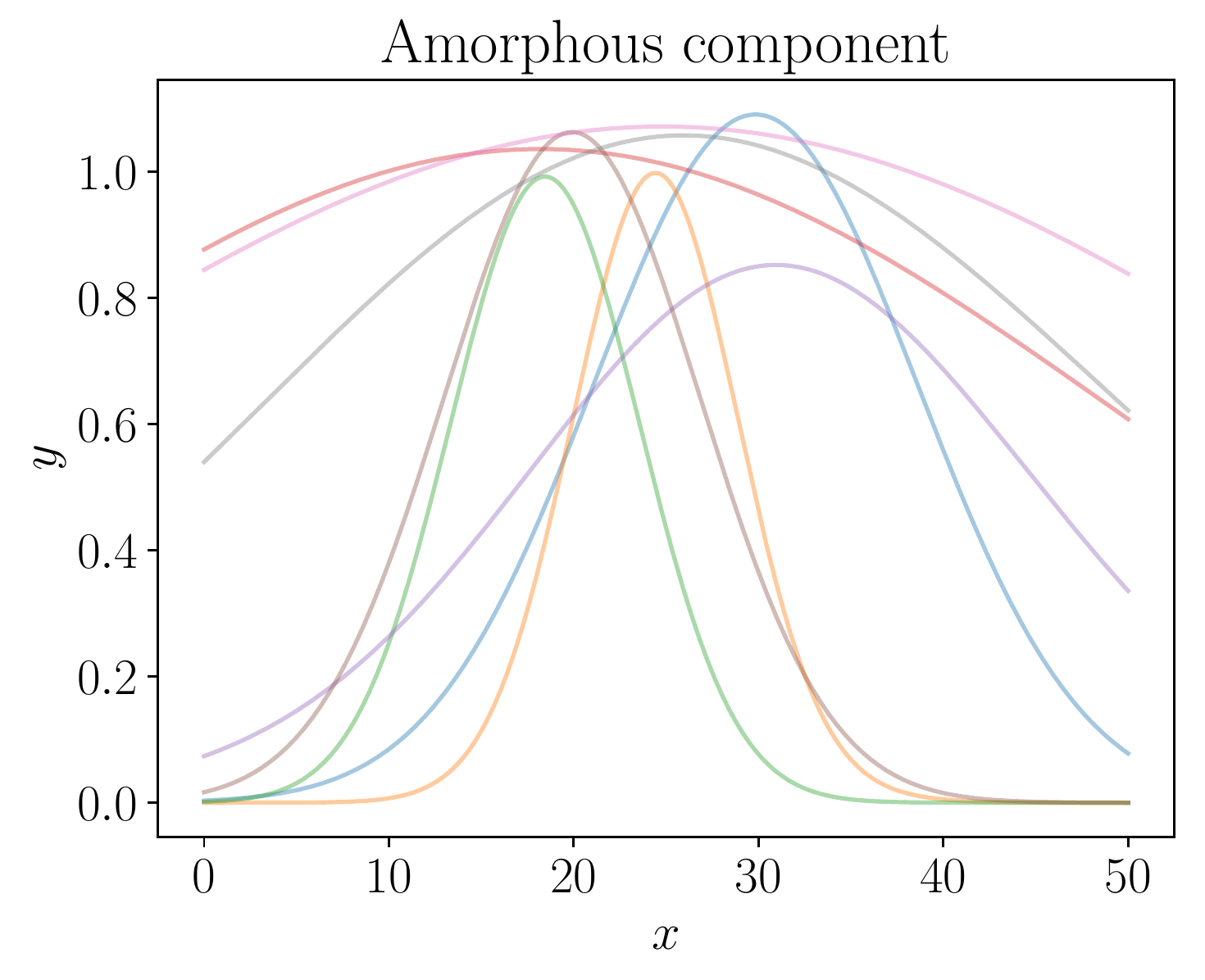}
\caption{\it Some examples of plausible shapes for the amorphous component,
generated from the prior distribution for its parameters
(Section~\ref{sec:priors}). For visual purposes only, these
were scaled to amplitudes $\sim$ 1. \label{fig:wide_component}}
\end{figure}

Figure~\ref{fig:wide_component} shows several examples of possible amorphous component shapes. The prior distributions (specified in
Section~\ref{sec:priors})
constrain the central positions of these gaussians much
more strongly than the central positions of the narrow peaks (crystalline component). AMORPH is primarily able to distinguish the crystalline
{\changed and amorphous components due to the different priors for the widths
of the two different sets of peaks.}

\subsection{Derived properties}
Some derived properties of the amorphous component can be used to summarize its shape.
We have included {\changed the `width' (second central moment, analogous to the standard
deviation)}, `skewness', which describes asymmetry, and `nongaussianity'
which quantifies the departure from gaussianity. These are not the same
concept, as the amorphous component could be non-gaussian while still being symmetric
(and thus having zero skewness).

Defining a normalized version of the amorphous component as
\begin{align}
g(\x) &= \frac{f_{\rm amorph}(\x)}
             {\int_{-\infty}^\infty f_{\rm amorph}(\x') \, d\x'},
\end{align}
its center of mass is
\begin{align}
\textnormal{center} &= \int_{-\infty}^\infty x g(\x) \, d\x.
\end{align}

The width of the amorphous component
(its second central moment, perhaps better interpreted
as a half-width) is defined by
\begin{align}
\textnormal{width} &= \sqrt{\int_{-\infty}^\infty
                            g(\x) \left(\x - \textnormal{center}\right)^2
                            \, d\x},
\end{align}
and the skewness is
\begin{align}
\textnormal{skewness} &= \int_{-\infty}^\infty
                           g(\x) 
                           \left(
                             \frac{x - \textnormal{center}}{\textnormal{width}}
                           \right)^3
                         \, d\x.
\end{align}

\begin{figure}[!ht]
\centering
\includegraphics[width=0.6\textwidth]{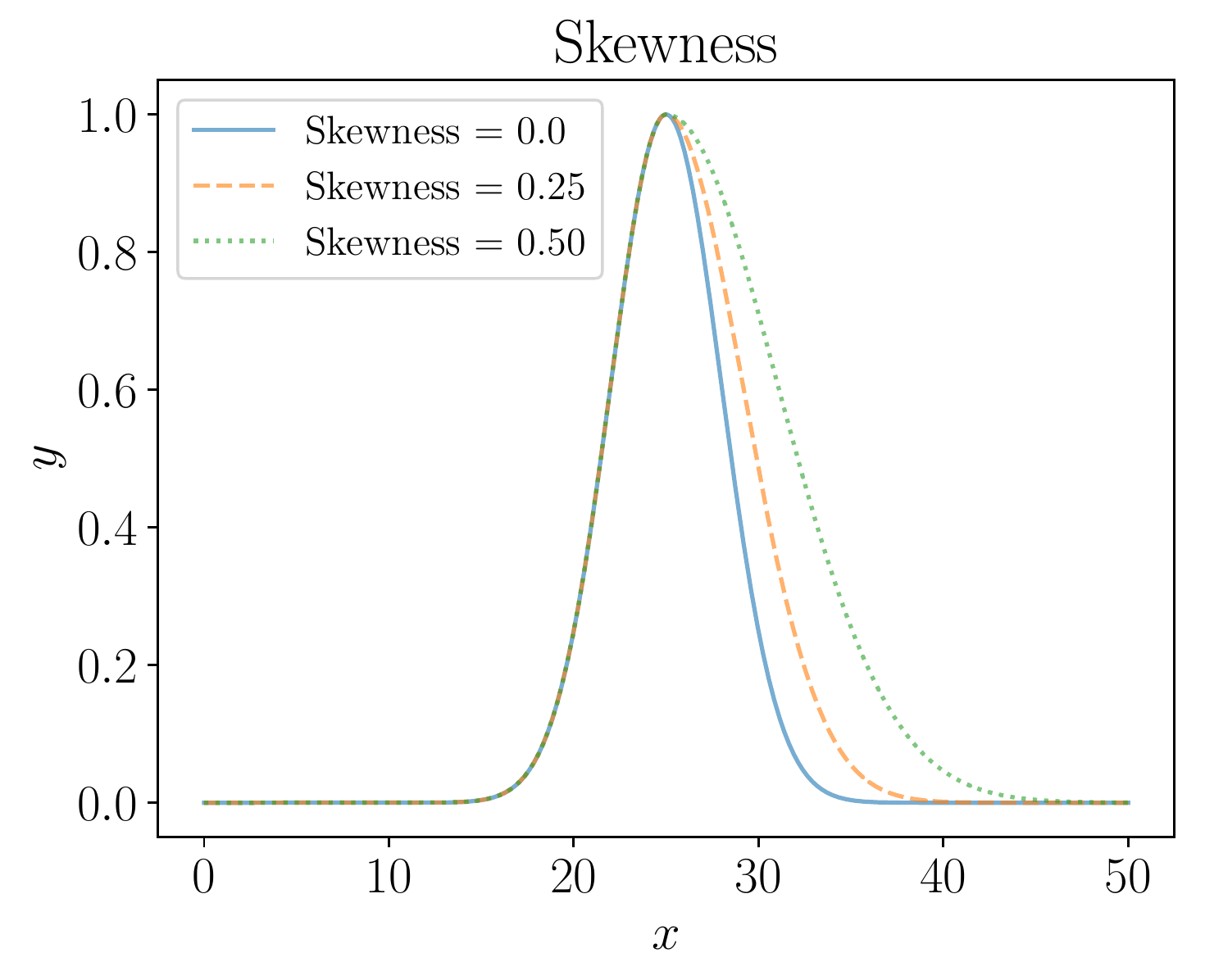}
\caption{\it Examples of amorphous component shapes with different values
for the skewness.\label{fig:skewness}}
\end{figure}

\begin{figure}[!ht]
\centering
\includegraphics[width=0.6\textwidth]{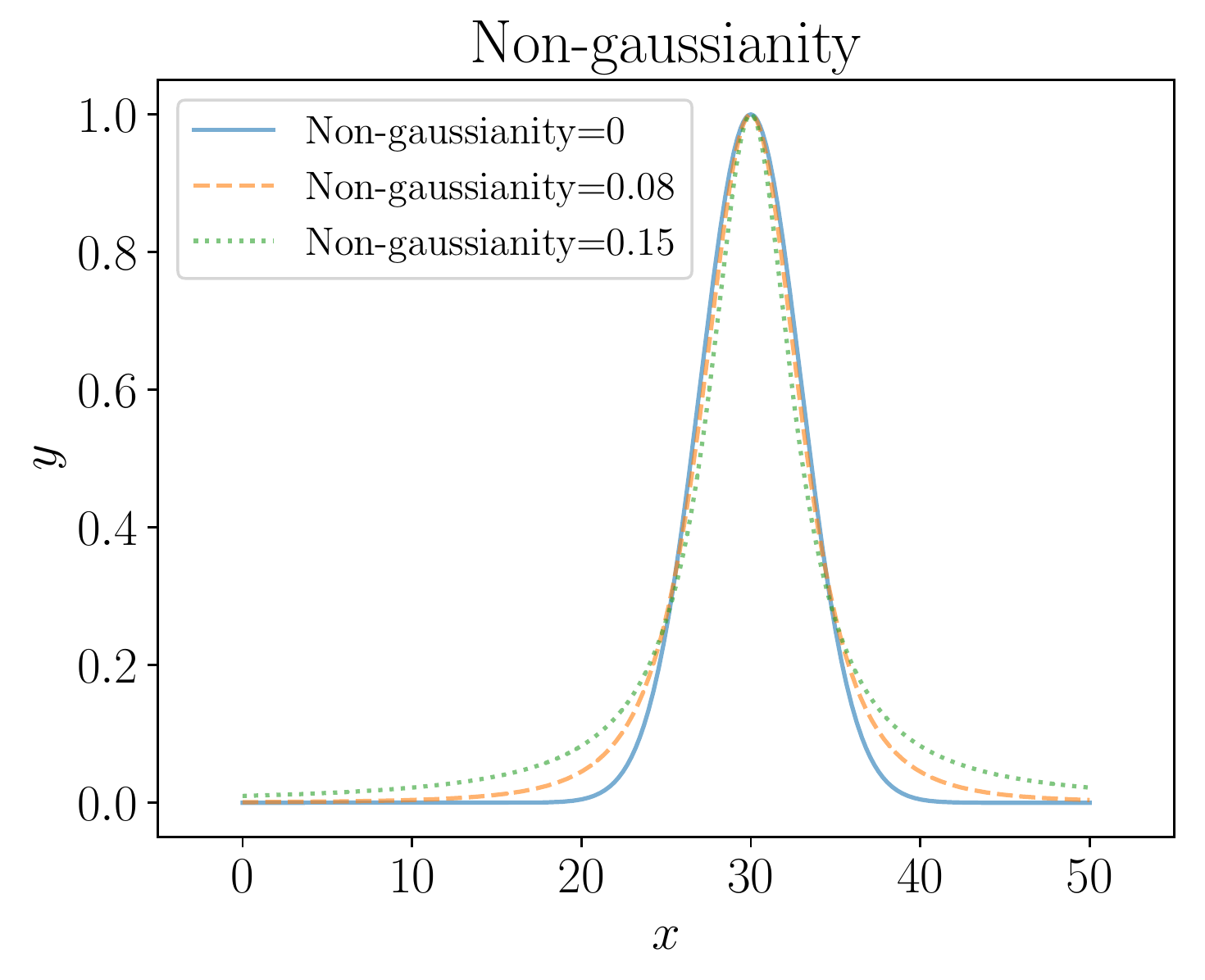}
\caption{\it Examples of amorphous component shapes with different values
for the nongaussianity.\label{fig:nongaussianity}}
\end{figure}

To quantify the non-gaussianity, we construct a gaussian
$g'(\x)$
with the same centre, width, and total integral as $g(\x)$, and compute the
Kullback-Leibler divergence from $g'$ to $g$:
\begin{align}
\textnormal{non-gaussianity} &= 
    \int_{-\infty}^\infty g(\x) \log\left[\frac{g(\x)}{g'(\x)}\right] \, d\x.
\end{align}
where
\begin{align}
g'(\x) &= \frac{1}{\textnormal{width} \times \sqrt{2\pi}}
            \exp\left[-\frac{1}{2\times (\textnormal{width})^2}
                    \left(\x - \textnormal{center}\right)^2\right].
\end{align}
The Kullback-Leibler divergence is a standard way of quantifying how
different one `measure' ($\equiv$ density function, for our purposes here) is from another \citep{knuth2012foundations}.
It is zero when the two measures are the same, in our case,
if $g(\x)$ is a gaussian. See Figures~\ref{fig:skewness} and~\ref{fig:nongaussianity} for example
shapes for the amorphous component, and corresponding values of the skewness
and nongaussianity.

\section{Prior probability distributions}\label{sec:priors}
The joint prior distribution for all of the unknown parameters and
the data must be specified.
When the number of parameters is large, this is often done hierarchically, by
specifying the priors conditional on some `hyperparameters', and then assigning
priors to the hyperparameters.

{\changed In these cases,
the joint prior distribution for
the hyperparameters $\hypers$}, parameters, and data,
is written $p(\hypers, \params, \data | \info)$, and is usually factorized
in this way:
\begin{align}
p(\hypers, \params, \data | \info) &=
    p(\hypers | \info)p(\params | \hypers, \info)
    p(\data | \params, \hypers, \info)\\
    &= p(\hypers | \info)p(\params | \hypers, \info)
    p(\data | \params, \info)
\end{align}
where the first step is true in general by the product rule
{\changed of probability}, and the second
step assumes that knowing the parameters would make the hyperparameters
irrelevant when predicting what data would be observed. In the AMORPH model,
we introduced hyperparameters describing the typical {\changed integrated area (`mass')}
and width of the peaks for the amorphous component, and
the degree to which the actual {\changed masses} and widths are scattered
around that typical value. This was also done for the crystalline
component.

\begin{figure}[!ht]
\centering
\includegraphics[scale=0.7]{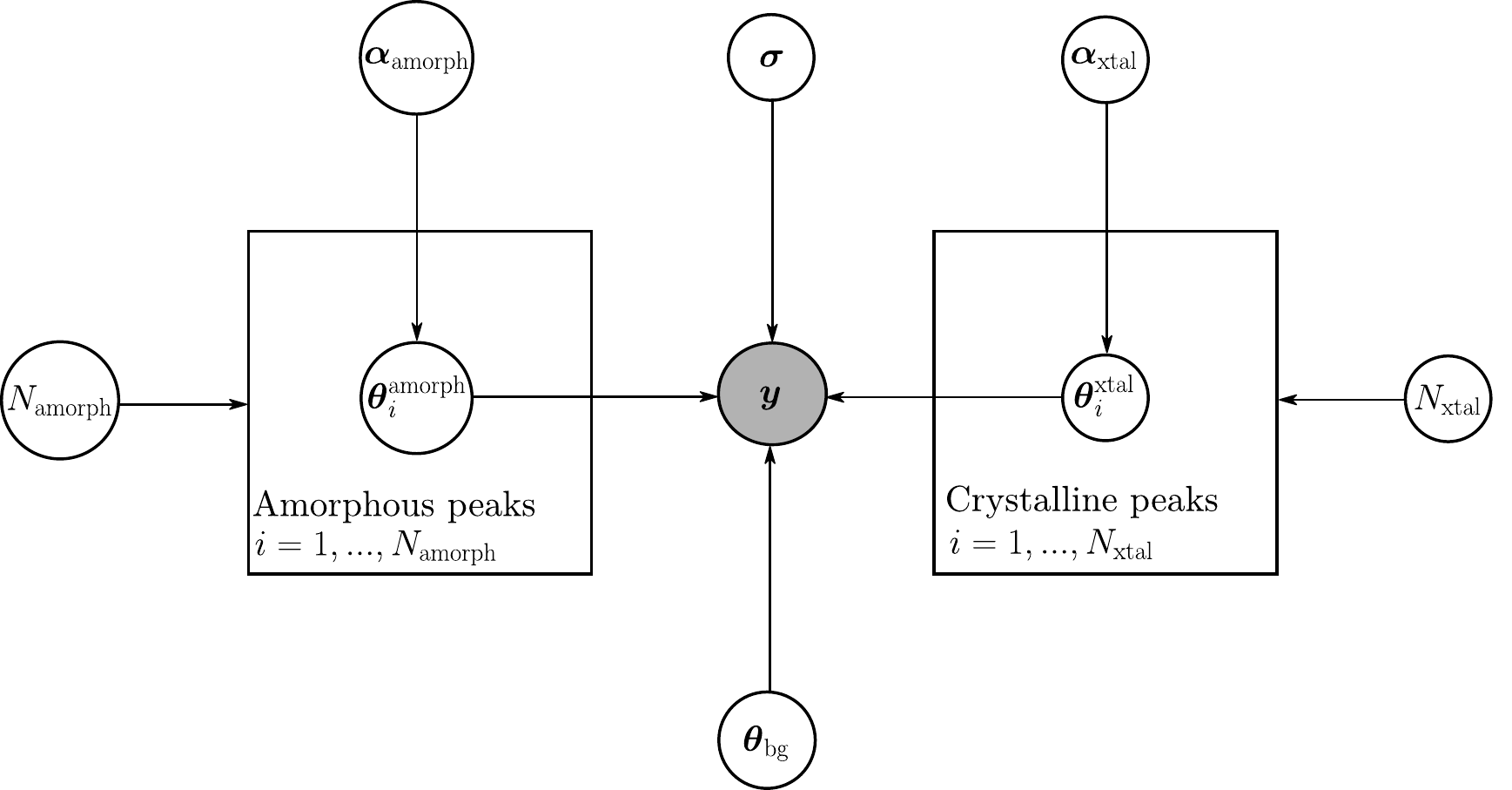}
\caption{The PGM. The hyperparameters controlling the prior distributions
for the amorphous and crystalline peaks are denoted collectively
by $\boldsymbol{\alpha}_{\rm amorph}$ and
$\boldsymbol{\alpha}_{\rm xtal}$ respectively.
The noise-related parameters are collected under $\boldsymbol{\sigma}$.
The data is $\boldsymbol{y}$.\label{fig:pgm-edited}}
\end{figure}

A directed acyclic graph, also known as a probabilistic graphical model (PGM),
which depicts the dependence structure of the entire joint
prior probability distribution,
is shown in Figure~\ref{fig:pgm-edited}. The prior distributions themselves
are specified in Table~\ref{tab:priors}.

\begin{table}[!ht]
\footnotesize
\centering
\begin{tabular}{|lll|}
\hline
{\bf Quantity}      &   {\bf Meaning}   &  {\bf Prior distribution}\\
\hline
{\em Prior information}&&\\
$n$ & Number of measurements & given\\
$\{\x_1, \x_2, ..., \x_n\}$  & $\x$-values of measurements & given \\
{\changed $x_{\rm range}$} & {\changed Range of measurements} & {\changed $\max(x_1, ..., x_n) - \min(x_1, ..., x_n)$} \\
\hline
{\em Hyperparameters} & &\\
$N_{\rm xtal}$   &   Number of narrow peaks    &  $\propto 1/(N_{\rm xtal}+1)$, $N_{\rm xtal} \in \{0, 1, ..., 1000\}$ \\
$N_{\rm amorph}$   &   Number of wide peaks    &  $\propto 1/(N_{\rm amorph}+1)$,
 {\changed $N_{\rm amorph} \in\{0, 1, ..., 1000\}$} \\
{\changed $a_{\ln M, \rm xtal}$} & Typical {\changed log-mass} of narrow peaks & Laplace(0, 5)\\
{\changed $a_{\ln M, \rm amorph}$} & Typical {\changed log-mass} of wide peaks & Laplace(0, 5)\\
{\changed $b_{\ln M, \rm xtal}$} & Diversity of {\changed log-mass} of narrow peaks & Uniform(0, 5)\\
{\changed $b_{\ln M, \rm amorph}$} & Diversity of {\changed log-mass} of wide peaks & Uniform(0, 5)\\
$a_{\ln W, \rm xtal}$ & Typical log-width of narrow peaks & LogUniform$(10^{-3}x_{\rm range}, 0.05x_{\rm range})$\\
$a_{\ln W, \rm amorph}$ & Typical log-width of wide peaks & LogUniform$(0.05x_{\rm range}, x_{\rm range})$\\
$b_{\ln W, \rm xtal}$ & Diversity of log-width of narrow peaks & Uniform(0, 0.1)\\
$b_{\ln W, \rm amorph}$ & Diversity of log-width of wide peaks & Uniform(0, 0.2)\\
\hline
{\em Background parameters}&&\\
$b$       & Mean background level       & $\ln b \sim $ Laplace$(0, 5)$\\
$\{n_1, ..., n_4\}$  & Background deviations & independent Normal$(0,1)$\\
\hline
{\em Crystalline parameters}&&\\
$X_{i, \rm xtal}$ & Positions of narrow peaks &
                            Uniform$(x_{\rm min}, x_{\rm max})$ \\
$M_{i, \rm xtal}$ & {\changed Masses} of narrow peaks &
{\changed
 $\ln M_{i, \rm xtal} \sim \textnormal{Laplace}\left(a_{\ln M, \rm xtal}, b_{\ln M, \rm xtal}\right)$} \\
$W_{i, \rm xtal}$ & Widths of narrow peaks &
 $\ln W_{i, \rm xtal} \sim \textnormal{Laplace}\left(a_{\ln W, \rm xtal}, b_{\ln W, \rm xtal}\right)$ \\
{\changed $\nu_{\rm xtal}$} &  {\changed Shape of narrow peaks (common)} &  {\changed LogUniform$(0.1, 100)$}\\
\hline
{\em Amorphous parameters}&&\\
$X_{i, \rm amorph}$ & Positions of wide peaks &
          Uniform$(x_{\rm min}+0.3x_{\rm range}, x_{\rm max}-0.3x_{\rm range})$ \\
{\changed $M_{i, \rm amorph}$} &  {\changed Masses} of wide peaks &
 {\changed  $\ln M_{i, \rm amorph} \sim \textnormal{Laplace}\left(a_{\ln M, \rm amorph}, b_{\ln M, \rm amorph}\right)$} \\
$W_{i, \rm amorph}$ & Widths of wide peaks &
 $\ln W_{i, \rm amorph} \sim \textnormal{Laplace}\left(a_{\ln W, \rm amorph}, b_{\ln W, \rm amorph}\right)$ \\
\hline
{\em Noise parameters}&&\\
$\sigma_0$ &    Constant noise level  &   $\ln \sigma_0 \sim \textnormal{Laplace}(0,5)$\\
$\sigma_1$ &    Noise proportional to $\sqrt{\textnormal{model}}$   &  $\ln \sigma_1 \sim \textnormal{Laplace}(0,5)$ \\
$\nu$     &   Shape parameter for $t$-distribution&LogUniform$(1, 1000)$\\ &for noise   &   \\
\hline
{\em Data}&&\\
$\{y_1, y_2, ..., y_n\}$  &   Measurements    & independent \\ && Student-$t\left(f(x_i; \params), \sqrt{\sigma_0^2 + \sigma_1f(x_i; \params)},\nu\right)$\\
\hline
\end{tabular}
\caption{\it The prior distributions for all hyperparameters,
parameters, and the data.\label{tab:priors}}
\end{table}

Most of the chosen distributions were intended to be reasonably
conservative and wide, {\changed but not excessively so}.
Some of the prior distributions in AMORPH use the log-uniform distribution,
sometimes (erroneously) called a Jeffreys prior, whose density for a quantity
$z$ is proportional to $1/z$. {\changed This prior is equivalent to a uniform prior
for the logarithm of the quantity, and encodes the fact that the order of magnitude
of a positive quantity may be unknown.}

We used a Laplace distribution for some informative priors, rather than
a more conventional normal distribution, because it assigns higher
prior probability to values far from the center  {\changed (compared to a normal
distribution)}, and can be computed conveniently. Specifically,
special functions such as the inverse of the error function are needed.
The Laplace distribution with location parameter
(central position) $a$ and scale parameter (width) $b$ has probability
density given by
\begin{align}
p(x | a, b) &= \frac{1}{2b}\exp\left[-\frac{|x - a|}{b}\right].
\end{align}

The priors for the positions and widths of the `amorphous' and `crystalline'
{\changed peaks} are different, to account for the prior knowledge that they
should result in wide {\changed peaks} near the centre of the domain, and
narrow {\changed peaks} throughout the domain respectively.

The prior for the data given all the parameters  {\changed (i.e., the sampling distribution)}
is a student-$t$ distribution,  {\changed centered at the model curve $f(x; \boldsymbol{\theta})$}.
{\changed The $t$ distribution} allows for potentially heavier tails than a normal distribution would.
Thus, analyses with AMORPH should be able to cope with discrepant datapoints
as long as they are not locally correlated (which would cause them to be
mistaken for peaks).

AMORPH does not assume a known number of {\changed peaks} for either the
amorphous or the crystalline component --- the appropriate number of
{\changed peaks} is inferred from the data. The maximum number of {\changed peaks}
for each component is set to {\changed 1000}.

\section{Applications}\label{sec:applications}
\subsection{Crystallinity}
{\changed To date, applications of crystallinity determinations via XRD methodologies on terrestrial samples have primarily focused on volcanic materials \citep[e.g.,][]{wall2014, ellis2015, andrade2017, zorn2018}, linking crystallinity of volcanic glass to timescales of geologic processes. The primary advantage of this technique being its ability to determine crystallinities for finer grain sizes than typically possible through standard petrologic techniques \citep{rowe2012}. In contrast, planetary science usages for determining crystallinity have largely focused on quantification of the amorphous and crystalline components for mass balance determination of physical and chemical components in diverse geologic materials \citep[e.g.,][]{blake2013, dehouck2014}}. 

\begin{figure}[!ht]
\centering
\includegraphics[width=1\textwidth]{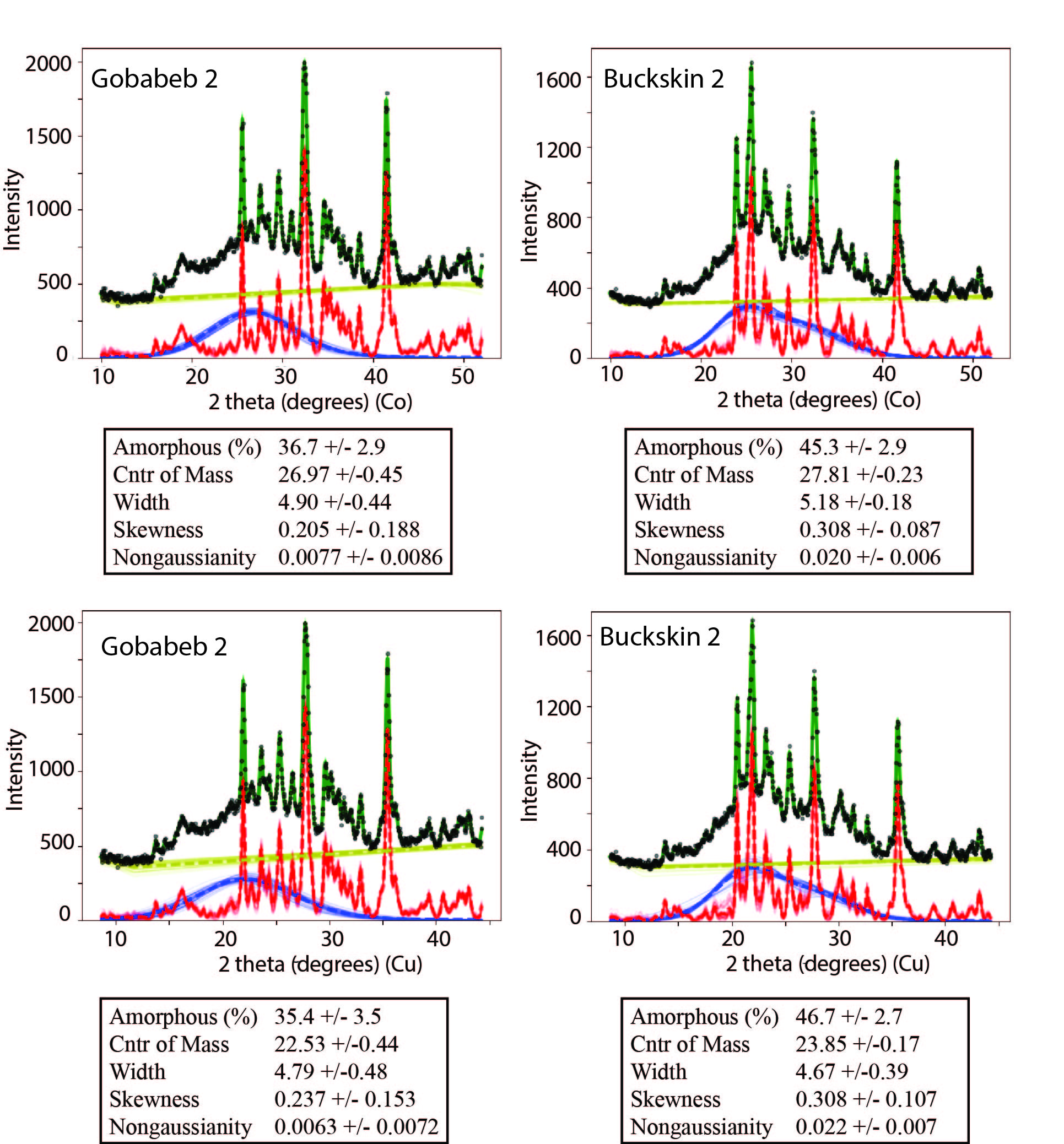}
\caption{\it \changed Analysis of Gobabeb 2 (left) and Buckskin 2 (right) results from the CheMin instrument on the MSL Curiosity rover (data sourced as above). Upper panels  illustrate initial Co K$\alpha$ processed results while lower panels show AMORPH results after translating data to a Cu K$\alpha$ source (see text for details). {\change Thin solid lines represent individual model results for crystalline (red) and amorphous (blue) components, linear backgrounds (yellow), and compiled model results (green). Bold dashed lines (same colors as individual model results) indicate average model results.} Calculated statistics provided by AMORPH for each analysis are indicated below the model results.\label{fig:Mars}}
\end{figure}

The AMORPH program calculates a fraction of amorphous material ($A_{\rm m}$) for each posterior sample of the parameters
using Equation~\ref{eqn:crystallinity}. Thus, measured crystallinity ($C_{\rm m}$) is easily determined from the relationship $C_{\rm m} = 1-A_{\rm m}$.
{\changed To test the applicability of AMORPH crystallinity results, we have analysed the Mars Scientific Laboratory (MSL)  CheMin instrument X-ray diffraction patterns of loose basaltic sand from the Bagnold Sand Dunes \citep[Gobabeb;][]{achilles2017, lapotre2017}
and sediments drilled from a silicic mudstone \citep[Buckskin sample;][]{morris2016}, both which have significant amorphous
materials (CheMin instrument data\footnote{\tt http://pds-geosciences.wustl.edu/msl/msl-m-chemin-4-rdr-v1/mslcmn{\textunderscore}1xxx.data.rdr4/}). CheMin diffraction patterns have been processed using the original Co K$\alpha$ ($1.7890$ \AA) data as well as data transformed from a Co K$\alpha$ to Cu K$\alpha$ ($1.5406$ \AA) X-ray source using Bragg's law. This recalculation simply allows for an easier visual comparison to terrestrial samples analysed with the more standard Cu K$\alpha$ X-ray source (see below). Note that for Co K$\alpha$ results, background critical points have been adjusted to 11.7--46.9\degree~$2\theta$ to maintain the linear background over the amorphous region from $\sim$ 8.8 to 2.25 \AA (Table ~\ref{tab:anode}). Although the results are uncalibrated, measured crystallinites provide a good first approximation of actual crystallinity, and relative differences are robust. Measured amorphous content of the Gobabeb 2 sample is 36.7 $\pm$ 2.9$\%$ (63.3$\%$ crystallinity) while the Buckskin 2 sediment contains 45.3 $\pm$ 2.9$\%$ amorphous phases (54.7 $\%$ crystallinity; Fig.~\ref{fig:Mars}). Comparatively, published FULLPAT calculations of amorphous contents are ~35 $\pm$ 15$\%$ for Gobabeb and 50 $\pm$ 15$\%$ for Buckskin \citep{achilles2017, morris2016}. }

{\changed The measured crystallinity is not equivalent to an ``actual'' or ``true'' crystallinity, although the values are often very close to one another as noted above.
Therefore,} this procedure does not negate the need for calibration standards
to translate measured crystallinity to an actual crystallinity.
{\changed Calibration standards can be easily created by mixing
powdered minerals (e.g. olivine, pyroxene, plagioclase) with powdered glass in known
proportions to simulate volcanic material of varying crystallinities \citep[e.g.,][]{rowe2012, wall2014}.} 

\begin{figure}[!ht]
\centering
\includegraphics[width=0.7\textwidth]{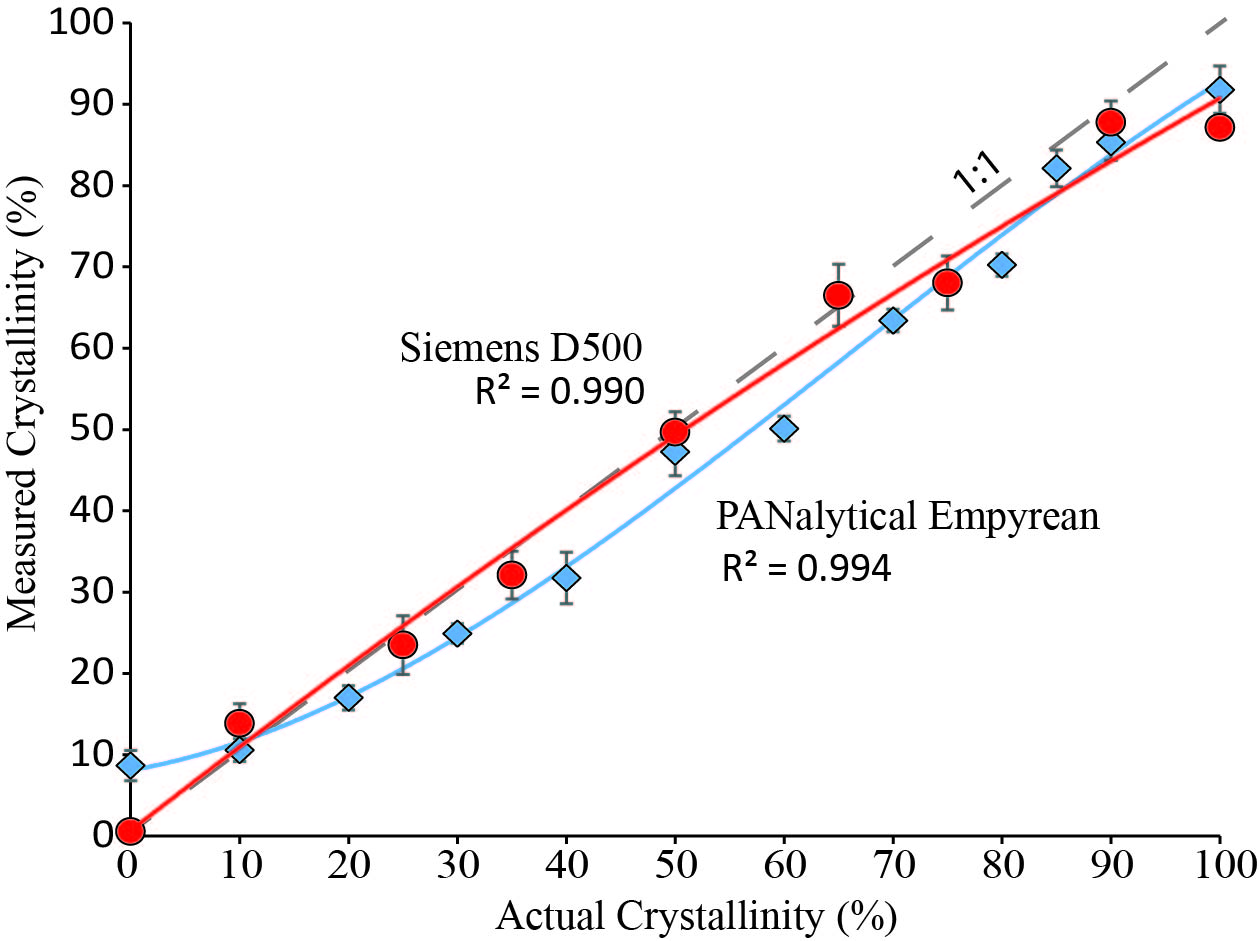}
\caption{\it Two calibration curves of the same standard material (rhyolite) are compared from different instrumentation, including a Siemens D500 (Washington State University; red circles) and a PANalytical Empyrean (University of Auckland; blue diamonds).\label{fig:instruments}}
\end{figure}

Transmission and absorption of X-rays in the material will vary as the ratio of mineral to glass changes in the sample, causing a deviation from a 1:1 line (Fig.~\ref{fig:instruments}). In addition to theoretical considerations, practical challenges also exist for background fitting. At low crystallinity ($\lesssim 10\%$) diffraction patterns are dominated by the low amplitude amorphous peak such that removing the effects of background noise becomes problematic, resulting in overestimated crystallinities. At high crystallinity ($\gtrsim 90\%$) the problem arises that peak overlaps in multi-mineral assemblages cause {\changed apparent} broadening effects, and may be subsequently modeled as ``amorphous'' and thus crystallinities  underestimated. The calibration curve therefore is utilized to remove these effects. This is also an important step as
different X-ray diffractometers will provide slightly different calibration curves (Fig.~\ref{fig:instruments}).
Differences between instruments largely reflect changes in background intensity and shape,
particularly at low values of $2\theta$.

\begin{figure}[!ht]
\centering
\includegraphics[width=1\textwidth]{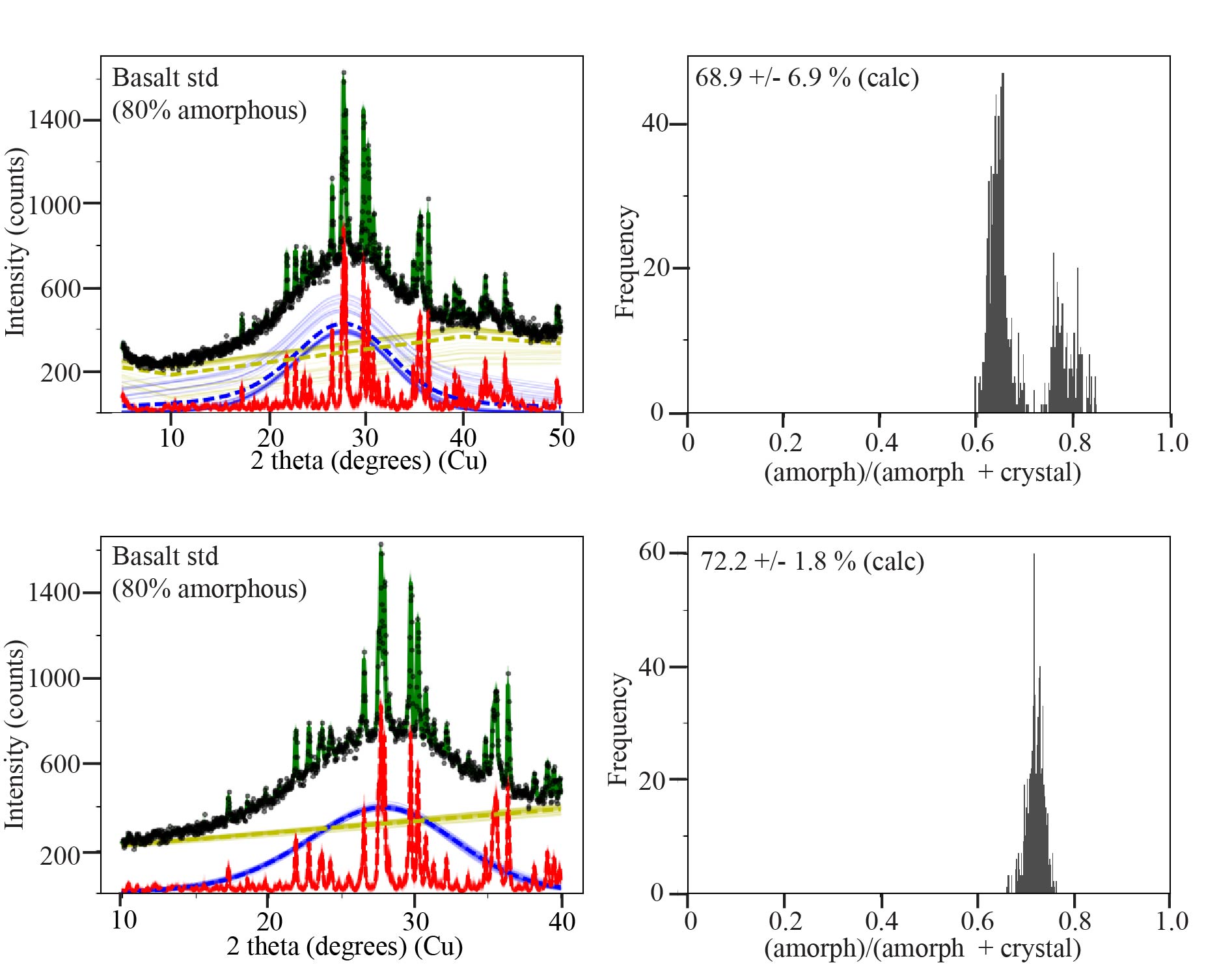}
\caption{\it \changed Model results for a basalt standard containing 80\% amorphous glass. Data was processed from 5--50\degree~$2\theta$ (upper left panel) with a calculated amorphous content of 68.9\% (upper right panel). The same data was processed instead from 10--40\degree~$2\theta$ (lower left panel) with a calculated amorphous content of 72.2\% (lower right panel). {\change Line colors same as in Figure~\ref{fig:Mars}.} The reasons for the
differences are discussed in the text.\label{fig:Outputs}}
\end{figure}

\begin{figure}[!ht]
  \begin{minipage}{\textwidth}
    \centering
    \includegraphics[width=0.85\textwidth]{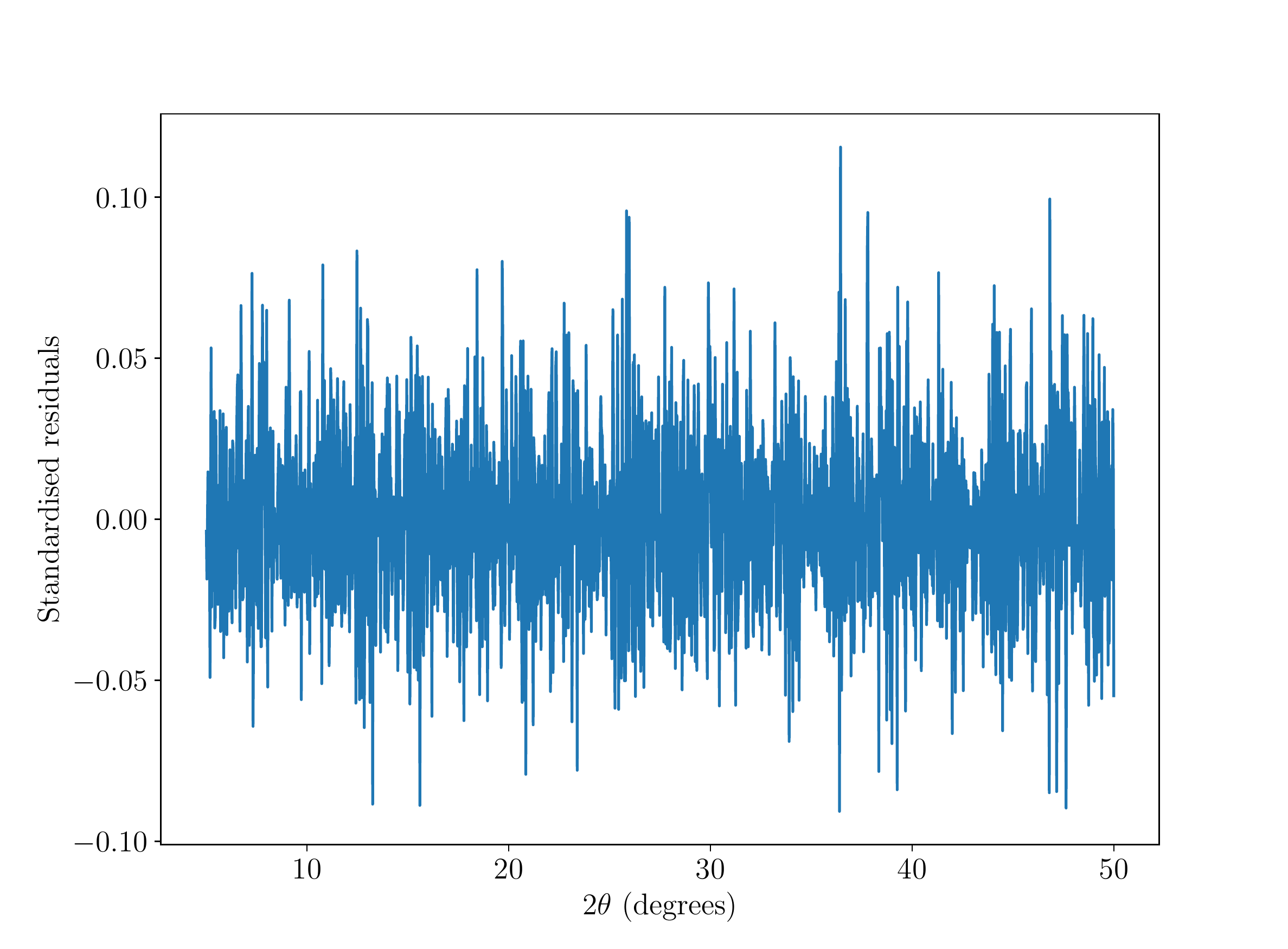} \\
    \includegraphics[width=0.85\textwidth]{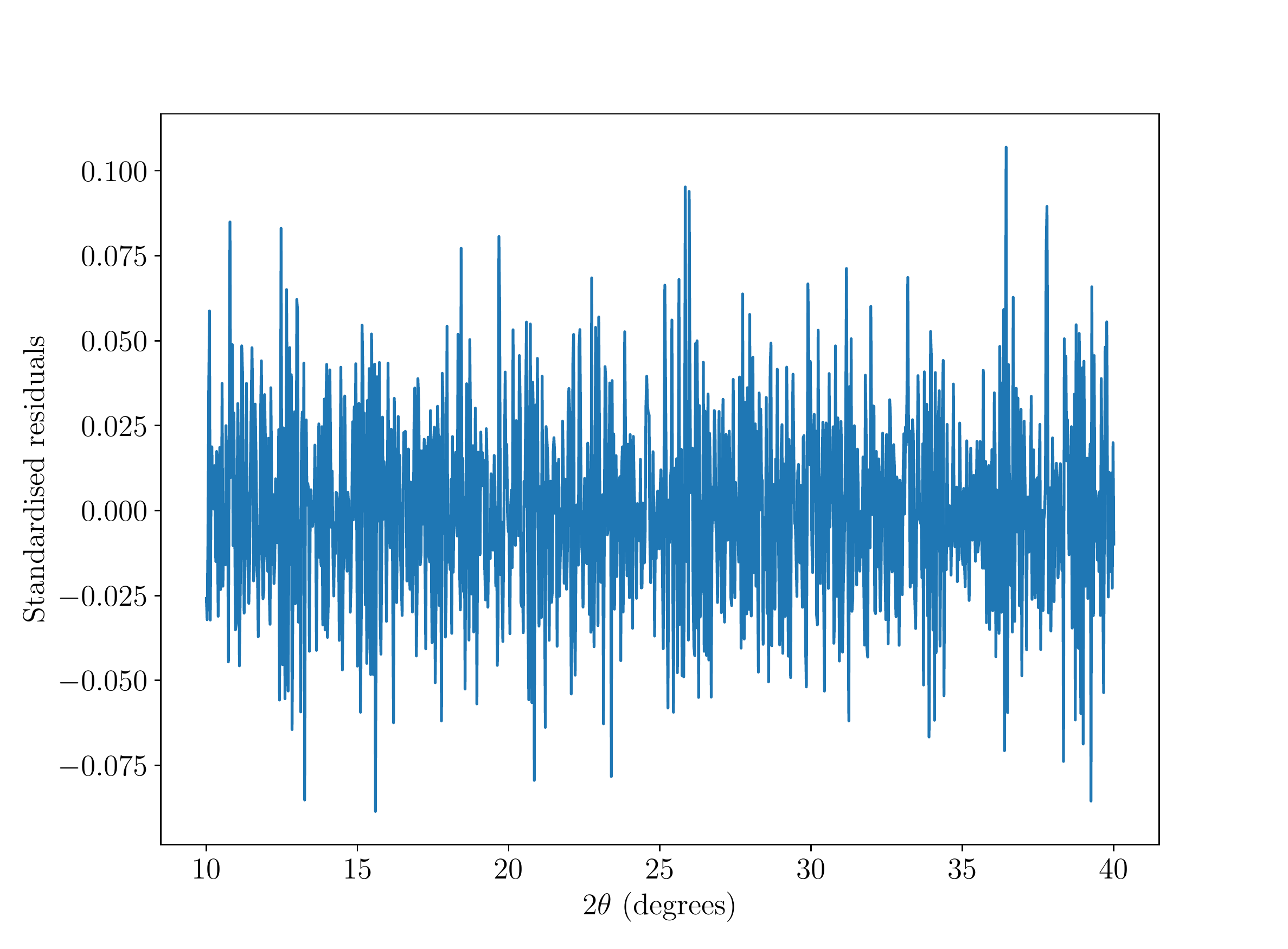}
  \end{minipage}
\caption{\it \changed Residuals of the model fits shown in Figure~\ref{fig:Outputs}, showing that the features of the data have been captured (upper panel is 5--50\degree~$2\theta$; lower panel is 10--40\degree~$2\theta$ dataset). Technically, there is no single set of
residuals, but a posterior distribution over them, represented by
the residuals of each curve in the generated sample. Therefore, these
plots show the residuals of the posterior mean model curve.\label{fig:residuals_etc}}
\end{figure}

\begin{figure}[!ht]
  \begin{minipage}{\textwidth}
    \centering
    \includegraphics[width=0.85\textwidth]{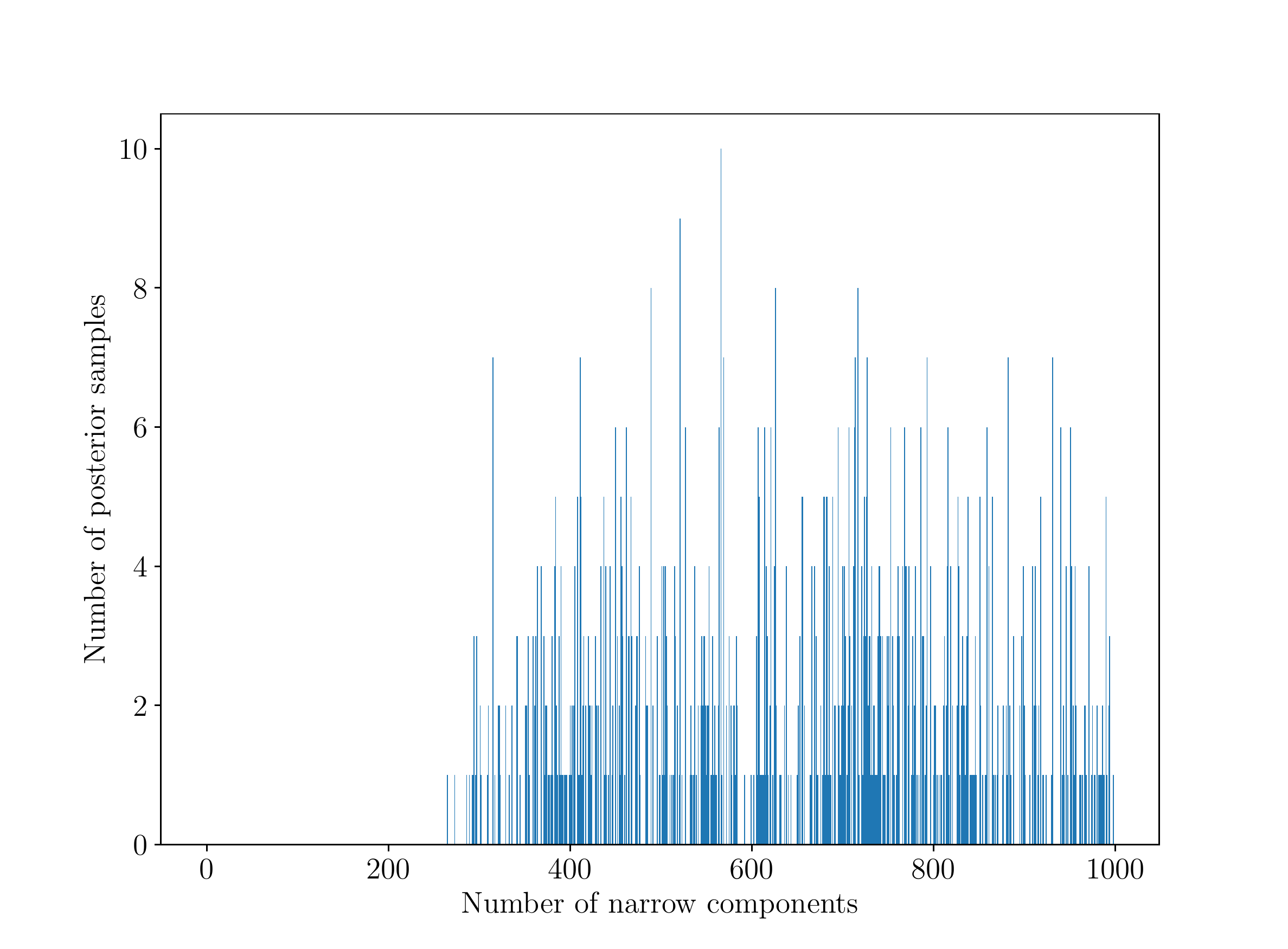} \\
    \includegraphics[width=0.85\textwidth]{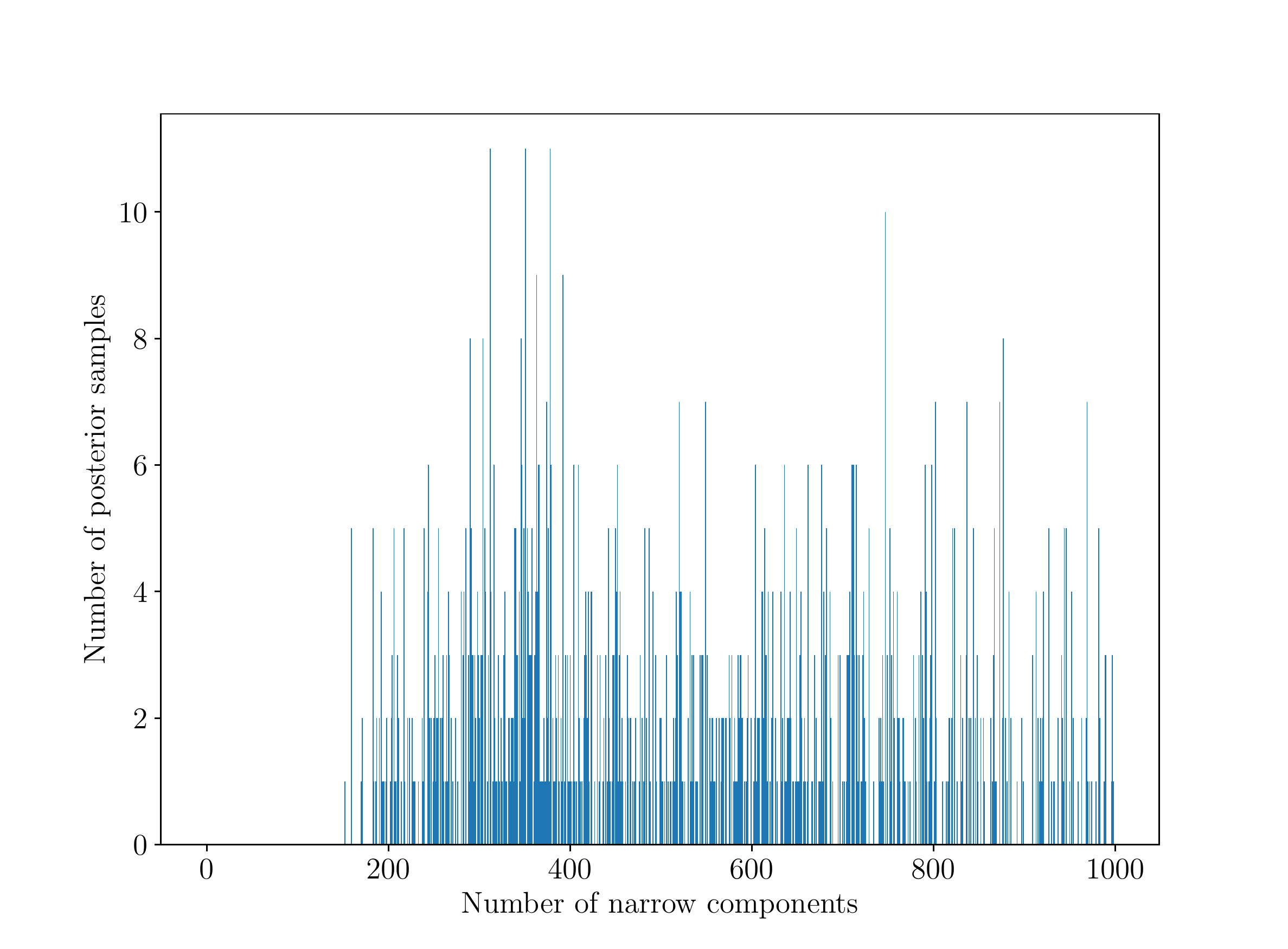}
  \end{minipage}
\caption{\it \changed Posterior distributions for the number of narrow peaks for the fits shown in Figure~\ref{fig:Outputs} (upper panel is 5--50\degree~$2\theta$; lower panel is 10--40\degree~$2\theta$ dataset). A minimum number of peaks is required in both cases, but the data cannot rule out large numbers of peaks (of very small amplitude). There are more peaks in the
5--50\degree case.\label{fig:num_peaks}}
\end{figure}

\begin{figure}[!ht]
\centering
\includegraphics[width=0.7\textwidth]{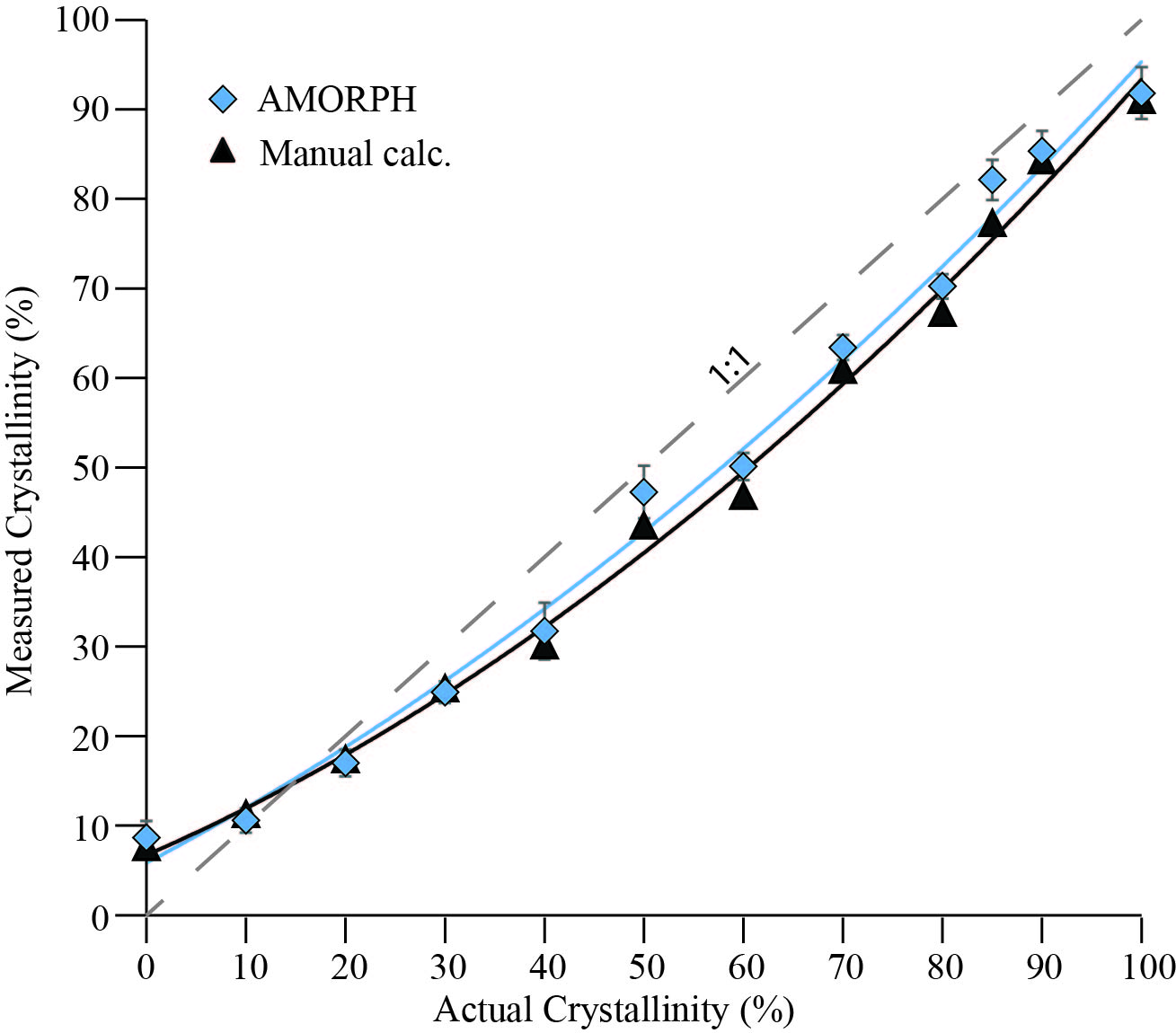}
\caption{\it \changed Comparison of calibration results for rhyolite calculated from AMORPH compared to a calibration calculated using the manual procedure described by \citet{rowe2012}.\label{fig:Comparison}}
\end{figure}

{\changed Since the wide peak for many} amorphous materials does not extend
appreciably out of the range from 10--40\degree~$2\theta$ (for a Cu K$\alpha$ source), peak {\changed area} integration
outside of this range will decrease {\changed proportionally} the relative contribution of the amorphous material
(only peaks for crystalline materials are outside this range {\changed in the example depicted in Figure~\ref{fig:Outputs}, with the corresponding fitting residuals and the posterior distributions
for the number of narrow peaks in Figure~\ref{fig:residuals_etc} and Figure~\ref{fig:num_peaks}, respectively).
It should be noted of course that this only effects the measured crystallinity and as long as the calibration standards are processed under the same conditions as the unknown, the ``actual'' crystallinity is not significantly affected by the choice in selected analysis interval. Results of the AMORPH calculations show good agreement to calibration curves using the manual
calibration approach for the same set of standards analyzed on the same PANalytical Empyrean X-ray diffractometer (Fig.~\ref{fig:Comparison}).}

{\changed For some materials, e.g. allophane, the amorphous region extends beyond the 10--40\degree~$2\theta$ range and thus would require analysis over a larger $2\theta$ interval. 
The trade-off in extending the $2\theta$ range is increased processing time and further deviation from a 1:1 calibration line. Since the analysis time is, in part, a function of the number of data points, if greater range in $2\theta$ is required, decreasing the total number of points (larger step size during analysis or manual removal of points in $2\theta$ regions not of interest) will offset additional analysis time. Thus we recommend that the optimal range for data processing in terms of time and accuracy is from 8.8 to 2.2 {\AA} or 10--40\degree~$2\theta$ for a Cu K{$\alpha$} X-ray source (see Table~\ref{tab:anode} for other X-ray sources) for crystallinity measurements of most volcanics. Importantly, reducing the data range from 5--50\degree~$2\theta$ to 10--40\degree~$2\theta$, does not increase the error associated with the peak fitting computational process with respect to other sources of error in sample selection and processing. In fact, for some analyses the variance in calculated crystallinity is appreciably reduced by shortening the range in analysed $2\theta$ and thereby reducing potential error associated with modeling background positions (Fig ~\ref{fig:Outputs}).}

\subsection{Amorphous Characteristics}
\begin{figure}[!ht]
\centering
\includegraphics[width=1\textwidth]{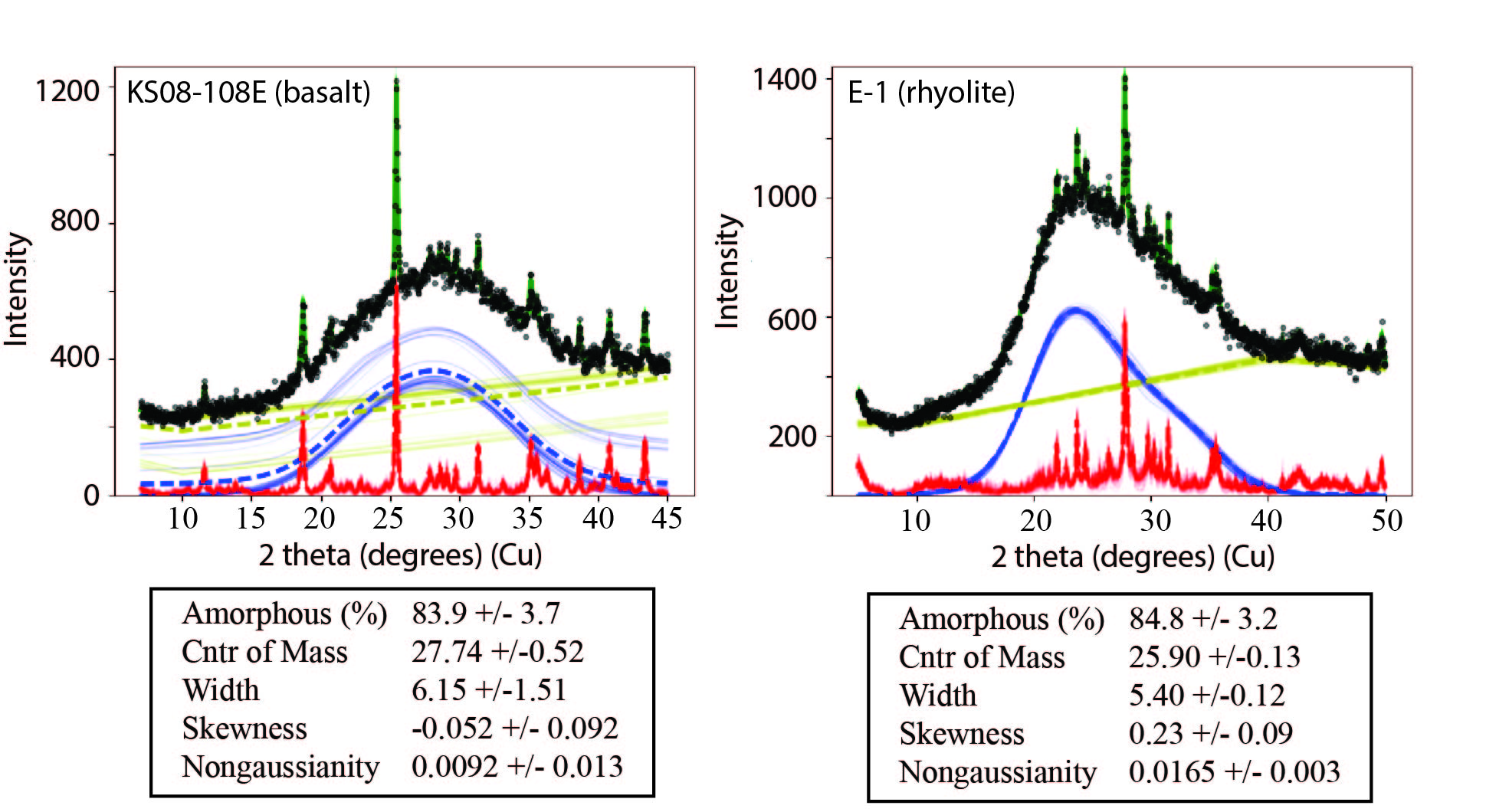}
\caption{\it \changed Examples of changing characteristics of amorphous material based on composition. Kilauea basalt glass (left) and taupo rhyolite glass (right) were analyzed on a PANalytical Empyrean XRD at University of Auckland. Calculated statistics for each analysis are provided below the model results. {\change Line colors same as described in Figure~\ref{fig:Mars}.} Note in particular the change in skewness from basalt to rhyolite.\label{fig:terrestrial}}
\end{figure}

As discussed above, one of the primary advantages of the AMORPH program is that it can
independently model characteristics of the amorphous component.  In particular, centre of mass,
skewness, and nongaussianity are distinctive amongst different amorphous materials. {\changed In contrast to crystallinity determinations, the focus of applications looking at the characteristics of the amorphous material are for indentification of the amorphous phase(s) \citep[e.g.,][]{dehouck2014}. To test
the ability of AMORPH to characterise amorphous phases}, we present amorphous characteristics of rhyolite and basalt glass picked from a
Taupo pumice \citep[73.5 wt\% SiO\textsubscript{2}; P2166C;][]{barker2015} and a Kilauea basalt
\citep[51 wt\% SiO\textsubscript{2}; KS08-108E;][]{wooten2009, rowe2015}, respectively.
{\changed The most notable differences in the amorphous component characteristics calculated for diffraction patterns for basalt and rhyolite glass include a positive skewness and shift to lower centre of mass
$2\theta$ values for the rhyolite glass compared to the basalt glass (Fig~\ref{fig:terrestrial}).} 

Although CheMin
diffraction analyses were run on a different X-ray diffractometer from the terrestrial
samples shown here, similar relative changes in amorphous characteristics can be observed
in X-ray diffraction data from Mars {\changed (Fig ~\ref{fig:Mars}). It is unknown if/how the X-ray source (e.g. Cu versus Co) fundamentally changes the peak shape characteristics of the amorphous content in this study. However, qualitative assessment suggests similar overall peak shapes between terrestrial samples analysed with a Cu X-ray anode (Fig.~\ref{fig:terrestrial}) and those analysed with a Co X-ray anode (Fig.~\ref{fig:Mars})}. In particular, the Buckskin 2 analysis of the amorphous component indicates a greater positive skewness {\changed than the Gobabeb 2 sample. In addition, the nongaussianity of the Buckskin 2 sample is $\sim 3\times$ greater compared to the Gobabeb 2 amorphous component}. These results suggest a change in the
composition of the amorphous material {\changed between the two Martian samples}. This interpretation, is consistent with prior
published results which suggest that at the Gobabeb locality (Namib Dune, Bagnold Dune Field),
the sample is dominated by a basaltic mineralogy, while at the Buckskin locality,
the amorphous component has been calculated to contain ~77 wt\% SiO\textsubscript{2}
\citep[i.e. rhyolitic;][]{morris2016, achilles2017}.

{\changed While results may imply a qualitative relationship to composition, further testing and an understanding of the program's limitations is necessary to better quantify compositional correlations. In particular, the AMORPH program can quantify the peak shape and proportion of amorphous material but it cannot deconvolve the diffraction pattern of multiple amorphous phases, nor can it distinguish between non-crystalline materials and X-ray amorphous phases such as nanophase oxides \citep{blake2013}.} 

\section{Conclusions}\label{sec:conclusions}
The AMORPH program’s statistical approach to interpreting amorphous materials in X-ray
diffraction patterns provides a new and unique methodology.
By explicitly specifying a set of assumptions and computing their
implications, we can automatically produce outputs separating the
amorphous and crystalline components and quantifying their properties,
along with a corresponding uncertainty estimate for any inferred
quantity.
A significant outcome of this research is that it reduces intra- and inter-user variability
by eliminating the need for manual background fitting, the largest source of error in the
calibration methodology for quantifying the amorphous component. Results demonstrate that
this approach 1) accurately reproduces values of known crystallinity and 2) is consistent
with prior manual approaches to the calibration method. Quantification of the amorphous
content however still requires a calibration curve to correct for x-ray absorption/emission
in heterogeneous materials, and to remove systematic biases in both background fitting and instrumentation. 

In addition to the quantification of amorphous and crystalline components, the AMORPH
program calculates statistical parameters of the amorphous component, including the centre
of mass, width of amorphous component, skewness, and nongaussianity. Characterization of the amorphous component requires no calibration and outputs show 
clear distinctions, in particular in terms of skewness, as a function of changing composition of the amorphous material in the demonstrated examples of
geologic materials from Earth and Mars.

\section*{Acknowledgments}

BJB was supported by a Marsden Fast-Start grant from the Royal Society of
New Zealand from 2013 to 2016, and this work is a spinoff from that
project. Data used in Figure~\ref{fig:Mars} analyzed and processed by J. Li, M. Jugo, A. Serrano. Manual calibration for Figure~\ref{fig:Comparison} was conducted by Y. Heled. This project made use of the resources of the Centre for
eResearch at the University of Auckland. {\changed We would like to thank E. Rampe and two anonymous reviewers for their comments.}

\appendix
\section{Installation and usage}\label{sec:program}

The entire repository, including C++ and Python source code, and a 
precompiled executable file for Microsoft Windows,
is hosted on Bitbucket at the following URL:

\vspace{1em}
{\tt https://bitbucket.org/eggplantbren/amorph}
\vspace{1em}

AMORPH is free software, released under the terms of the GNU General Public
Licence, version 3.
For most users, we suggest simply using the pre-compiled Windows executable
{\tt AMORPH.exe} from the repository. {\change Before executing the program, open
{\tt config.yaml} in a text editor to set up the run. This file contains all the
information about the run, such as the name of the file containing the data.
Then run {\tt AMORPH.exe}.} All files to be processed by AMORPH need to be in .txt file format, space delimited, with no headers. For simplicity, text files for processing should be located in the same folder as the {\tt AMORPH.exe} program.

{\change 
The configuration file lets you control the data being analyzed, and
various options such as the number of simultaneous threads to run.
You can also choose to use either
{\tt OPTIONS\_AGGRESSIVE} or {\tt OPTIONS} (or your own file),
which contain parameter settings for DNest4.
{\tt OPTIONS} is more conservative than {\tt OPTIONS\_AGGRESSIVE},
as the names suggest.
}

The program is set to run until 10,000 saved parameter sets
{\changed (`particles')} have been generated. Data outputs may be viewed at any point, however, closing the program before reaching 10,000 will reduce the accuracy of the final calculations. After running for a while, the output can be viewed
by running the Python script {\tt showresults.py}:

\vspace{1em}
{\change {\tt python3 showresults.py}}
\vspace{1em}

This script makes use of the packages {\tt Numpy} and {\tt matplotlib},
and has only been tested under Python 3. Anaconda\footnote{\tt https://www.anaconda.com/download/} is a convenient distribution
of Python which comes with scientific packages pre-installed.
Since AMORPH is really just a specific data analysis situation implemented
for DNest4 \citep{dnest4}, that paper provides much more detail about
AMORPH's output and further available options.

\end{document}